\documentstyle[aps,prl,twocolumn,graphicx,epsfig]{revtex}  

\draft
\begin{document}

\twocolumn[\hsize\textwidth\columnwidth\hsize\csname
@twocolumnfalse\endcsname

\title{Mg-Ni-H films as selective coatings: tunable reflectance by layered
hydrogenation}
\author{J.L.M. van Mechelen, B. Noheda, W. Lohstroh, R.J. Westerwaal, J.H.
Rector, B. Dam and R. Griessen}
\address{Condensed Matter Physics, Vrije Universiteit, De Boelelaan 1081, 1081 HV
Amsterdam , The Netherlands}

\maketitle

\begin{abstract}
Unlike other switchable mirrors, Mg$_{2}$NiH$_{x}$ films show large changes
in reflection that yield very low reflectance (high absorptance) at
different hydrogen contents, far before reaching the semiconducting state.
The resulting reflectance patterns are of interference origin, due to a
self-organized layered hydrogenation mechanism that starts at the substrate
interface, and can therefore be tuned by varying the film thickness. This
tunability, together with the high absorptance contrast observed between the
solar and the thermal energies, strongly suggests the use of these films in
smart coatings for solar applications.
\end{abstract}


\vskip2pc]

\narrowtext

\marginparwidth 2.7in
\marginparsep 0.5in

The discovery of the YH$_{x}$ switchable mirrors by Huiberts \textit{et al.}~%
\cite{Huiberts} in 1996 has been followed by extensive studies on rare-earth
hydride films. Upon absorbing hydrogen, a switchable mirror transforms from
a shiny metal to a transparent semiconductor. The same behaviour was later
observed in Mg-alloyed rare-earth hydrides~\cite{Sluis} and Mg-Ni hydrides~ %
\cite{Richardson}. However, at low hydrogen concentration the Mg-Ni hydrides
also present a third intriguing ''black'' state with low reflectance and
zero transmittance in the whole visible range\cite{Isidorsson}, which
immediately suggests their application as switchable smart coatings in, e.g.
solar heat collectors.

Since this type of applications generally involves absorption of the solar
spectrum (at photon energies between 0.5$<$ $E<$ 4 eV) and/or emission of
thermal radiation (at $E<$ 0.5 eV, for 100$^{o}$C)\cite{solar},
investigation of the switching behavior in the infrared is also essential.
In this letter, we study the peculiar optical patterns observed in Mg$_{y}$%
NiH$_{x}$ films ($y\sim $2) during hydrogenation (0$<x<$4) in the infrared
between $0.2<E<1$ eV ($1.2<\lambda <6.2$ $\mu $m), with special attention to
their applicability as selective tunable absorbers.

Films of Mg$_{y}$Ni (1.4$<y<$2.4), with a thickness between 140 and 500 nm,
are deposited on CaF$_{2}$ substrates by magnetron sputtering from a Mg
target with Ni additions. The background and Ar pressures are $10^{-5}$ Pa
and 1 Pa, respectively. The films are covered with a thin (3--11 nm) Pd
layer both to protect the films against oxidation and to catalyze the
hydrogen uptake. Profilometry and Rutherford backscattering (RBS) are used
to measure the film thickness and composition. Optical measurements at near
normal incidence ($\sim 15^{\circ }$) are performed from the substrate side 
\textit{in situ} during hydrogen loading and unloading in a Bruker IFS 66
Fourier transform infrared spectrometer. Simultaneously, the electrical
resistivity of the film is recorded in a Van der Pauw configuration\cite%
{Pauw}. During loading, the hydrogen gas pressure is adjusted between $%
10^{2} $ and $10^{5}$ Pa, depending on the kinetics of the sample. For
unloading, the films are exposed to air at temperatures between 30 and 120$%
^{\circ }$C.

Mg$_{y}$Ni ($y\sim 2$) films exhibit metallic behavior with resistivities, $%
\rho $, of about 50 $\mu \Omega $cm and high reflectance values that vary
between 0.85 at 0.2 eV and 0.6 at 1 eV. As shown in Fig. 1 for 0.85 eV ($%
\lambda $= 1.46 $\mu $m), already a small amount of H in the films (short
loading times or low $\rho $) is enough to dramatically decrease the
reflectance while the resistivity still shows clear metallic behavior. This
highly-absorbing state is the natural extension to infrared energies of the
''black state'' observed in Mg$_{2}$NiH$_{x}$ films in the whole solar range%
\cite{Isidorsson,Enache}. Further increase of the hydrogen content produces
alternating recovery and loss of reflectance and reveals that the
highly-absorbing states are not, or not only, related to a particular
hydrogen composition. Finally, Mg$_{2}$Ni is transformed into Mg$_{2}$NiH$%
_{4}$, which is a semiconductor\cite{Enache} with a band gap of about 1.6 eV%
\cite{Myers}, and the film becomes transparent, as indicated by the onset of
the transmittance $T$ in Fig. 1. In the fully loaded state $\rho $ is of the
order of 10 m$\Omega $cm\cite{note1}.

\begin{figure}
\begin{center} 
\includegraphics[width=8 cm,clip]{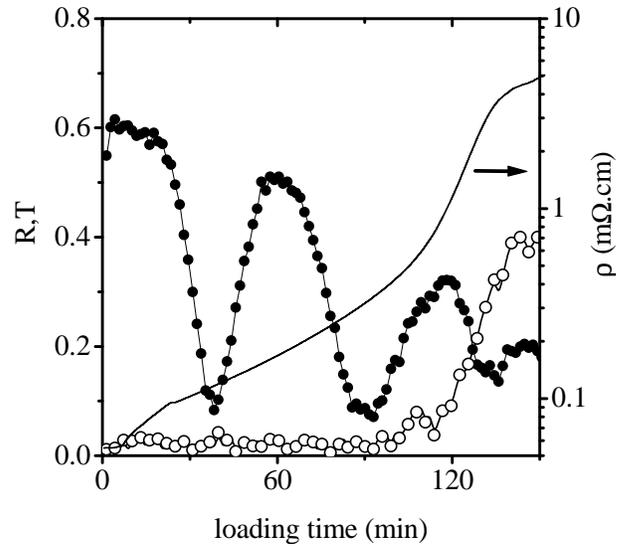} 
\end{center} 
\caption{{\protect\small Reflectance (}$\bullet ${\protect\small ),
transmitance (o) and resistivity (}$-${\protect\small ) as a function of
hydrogen loading time of a 420 nm Mg}$_{1.80}${\protect\small Ni film with a
3 nm Pd cap layer at E= 0.85 eV (}$\protect\lambda ${\protect\small = 1.46 }$%
\protect\mu ${\protect\small m). The reflectance is measured from the
substrate side of the sample.}}
\end{figure}

Figure 2 shows contour plots of the reflectance $R$, as a function of photon
energy, $E$, and resistivity (an indirect measure of $x$\cite{Enache}),
during loading, for films of various thickness, $d$. For a specific $E$, the 
$R$ pattern as a function of $\rho $ produces a plot similar to that in
Fig.1. In Fig. 2a the reflectance is also depicted during unloading, showing
the reversibility of the hydrogenation process. It is clearly observed that
the reflectivity oscillations during loading (or unloading) evolve into the
interference fringes of the fully-loaded transparent state, which already
points to a common origin for both of them. Since the measurements are
performed from the substrate side, the existence of interference already in
the first stages of the hydrogenation process, when there is still no
transmission through the film, can only be explained by the formation of a
well defined transparent layer, presumably Mg$_{2}$NiH$_{4}$, at the
substrate-film interface\cite{Lohstroh}. This layer, of thickness $t(x)<d$,
grows during H loading and finally reaches the total thickness of the film, $%
d$. The two main features observed in Fig 2, namely, the hyperbolic-like
bending of the minima and maxima as a function of the energy, and the
increment of the number of reflectance oscillations with increasing
thickness, are consistent with this model. This can be qualitatively
understood from the interference condition $2t(x)n=N\lambda $ $\propto N/E$,
where $n$ and $N$ are the refraction index and interference order,
respectively\cite{Born&Wolf}. Quantitative evidence of this phenomenon has
been obtained by modelling the reflection and transmission of a three-layer
stack (Pd-Mg$_{2}$NiH$_{0.3}$-Mg$_{2}$NiH$_{4}$) at various hydrogenation
stages\cite{Lohstroh}. It is worth to notice that the same self-organized
layered mechanism occurs during unloading in a completely reversible way
(see Fig. 2a).

The observed fringes lead to a switchable and highly tunable reflectance, $R$%
, or absorptance, $A$ (during loading, when there is no transmission through
the film, $A=1-R$), and strongly suggest the use of these films as smart
coatings. Figure 3 shows the reflectance as a function of photon energy for
different loading stages (A to E in Fig. 2a), in a film with $d$= 475 nm.
The metallic reflectance of the unloaded film (stage A), as high as 0.85 at
0.2 eV, is plotted in Fig. 3a. With only a small amount of hydrogen added,
at a composition of about Mg$_{2}$NiH$_{0.6}$ (the average H content
depending on $d$\cite{Lohstroh}), the film becomes black. The reflectance
decreases drastically at $E>$ 0.5 eV ($\lambda <$2.5 $\mu $m), and varies
between 0.05 and 0.3 at 0.5$<E<$ 4 eV (see Fig. 3b and ref.\cite%
{Isidorsson,Lohstroh}), while it is still reflecting ($R>$ 0.6) in the
thermal range. By linearly extrapolating the reflectance of Fig. 3b to low
energies and convoluting $A=1-R$ with the blackbody radiation at 100$^{o}$C,
one can estimate the thermal emittance in this state to be about 0.16.
Similarly, a solar absorptance of about 0.84 is estimated by using the data
at $E>$ 0.5 eV in the same figure together with those obtained for similar
films at energies up to 4 eV\cite{Isidorsson,Lohstroh}. These high solar
absorptance and low thermal emittance are comparable to those recently
reported for films of a-Si:H/Ti on Al substrates\cite{Schuler}, proposed as
good candidates for selective coatings in solar collectors. The main
advantage of the Mg$_{2}$Ni films is that their behaviour is switchable.
Moreover, the small amount of H needed, which involves minimal morphology
changes and thus little aging, makes the switching between A and B very
attractive from the applications point of view. Another advantage of these
films is the simplicity of the design since the active layer is a single
film which in itself switches between reflecting and absorbing.

\begin{figure}
\begin{center} 
\includegraphics[width=8 cm,clip]{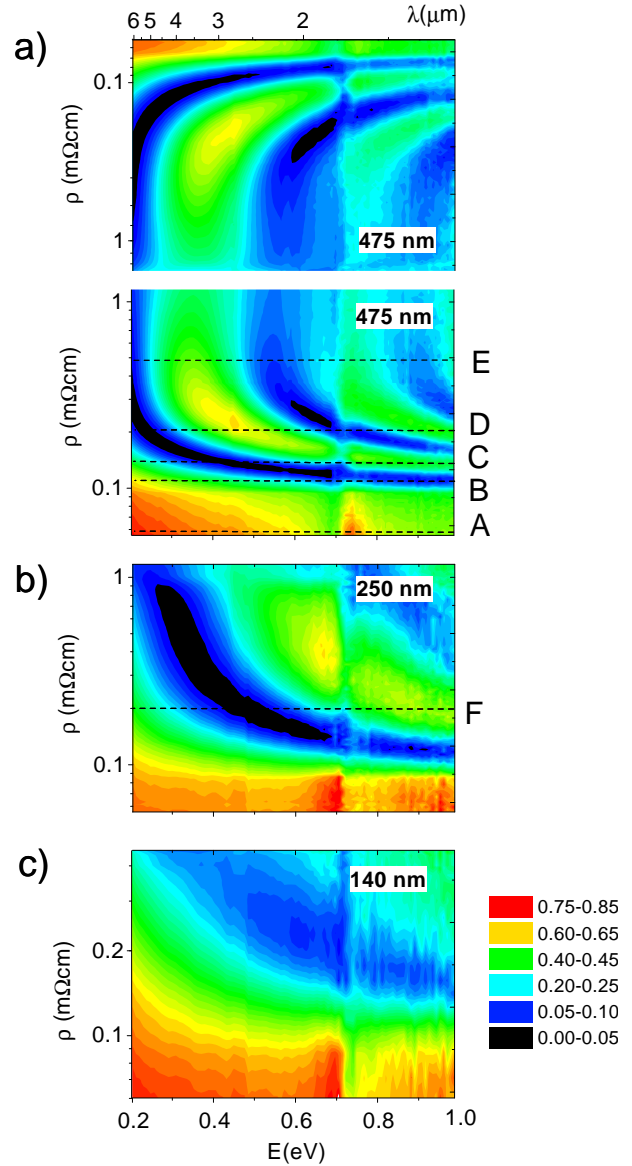} 
\end{center} 
\caption{{\protect\small Contour plots of reflectance vs. photon energy and }%
$\protect\rho ${\protect\small \ for a a) 475 nm Mg}$_{1.96}${\protect\small %
Ni,} {\protect\small \ b) 250 nm Mg}$_{1.70}${\protect\small Ni, and c) 140
nm Mg}$_{1.43}${\protect\small Ni film.}}
\end{figure}

Upon increasing the H content (stage C) a very different reflectance pattern
is measured with a narrow highly absorbing state at $\sim 0.4$ eV ($\lambda
\simeq $ 3.1 $\mu $m), whose energy and width can be tuned by varying the
film thickness. Due to the layered loading mechanism, the reflectance
spectrum at this state of the loading process is identical to that of a
later loading stage in a thinner 250 nm film (F), as shown in Fig. 3c. A
larger H content in a thick enough film gives raise to more than one $R$
minima (D).

As shown above, a transparent Mg$_{y}$NiH$_{x}$ ($x=4$ for $y=2$)\ growing
layer determines the reflectance spectra, even at the initial loading
states. The spectrum in Fig. 3d shows the interference pattern observed when
the whole film is transparent. The solid line is the result of fitting $R$
and $T$ ($T$ not shown for clarity) to those of a Mg$_{y}$NiH$_{x}$ film
with a Pd cap layer, both of known thickness and unknown $n$ and absorption
coefficient, $k.$\cite{Lohstroh} The optical constants obtained for the
transparent layer from the fits of the films under study are in between 2.5$%
\leq n\leq $4.0 and 0.01$\leq k\leq $1.5, depending on composition, in the
investigated energy range. The lowest $k$ values are desirable since they
produce the largest interference contrast. This can be observed by comparing
the high contrast of the film in Fig. 1, with $k\approx $ 0.03 and the lower
contrast of the film in Fig. 2c with $k\approx $ 1.

\begin{figure}
\begin{center} 
\includegraphics[width=7 cm,clip]{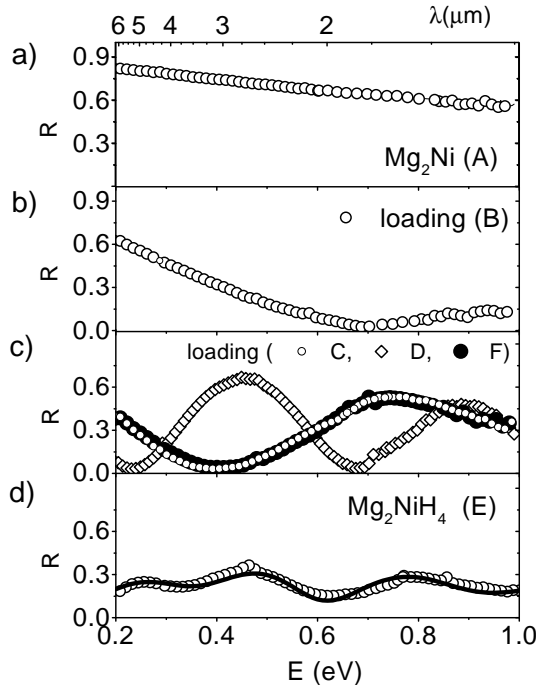} 
\end{center} 
\caption{{\protect\small Reflectance spectra for the 475 nm Mg}$_{1.96}$%
{\protect\small Ni film shown in Fig. 2c at different stages (A to E) of the
H loading. The letters refer to those in Fig. 2c. The solid circles in c)
correspond to the 250 nm Mg}$_{1.7}${\protect\small Ni film. The line in d)
is the fit to the data with n}$\approx ${\protect\small 3.7 and k}$\approx $%
{\protect\small \ 0.5, both weakly depending on energy in the studied range. 
}}
\end{figure}

Nucleation of the transparent layer at the film-substrate interface is
observed for all compositions under study, however the Mg/Ni ratio
influences the growth of the transparent layer. In slightly Mg-rich films
with 2.0$<y<$ 2.4 only the first minimum and maximum are observed, in
agreement with Isidorsson et al.\cite{Isidorsson}. The contrast is also
found to greatly decrease with the Mg content for 2.0$\leq $ $y$ $<$ 2.4,
pointing to the formation of a hydride layer with ill-defined interfaces.
Whether the role of the Mg/Ni ratio is direct, e.g. due to the catalytic
properties of Ni, or indirect, e.g. due to the low degree of crystallinity
of the Ni-rich films\cite{Crystal} is not clear yet. Work is in progress to
elucidate the catalytic properties of the substrate-film interface and the
role of the Mg/Ni ratio during loading.

In summary, Mg$_{y}$NiH$_{x}$ films with y$\approx $ 2 are particularly
suitable as smart coatings due to their energy tunability, high reflectance
contrast and switching capabilities. While the metallic state reflects about
60 \% of the incoming solar radiation, a small amount of H is enough to
change from reflective to black, absorbing 84\% of the solar spectrum. As
the film in this state emits only 16\% of the 100$^{o}$C blackbody radiation
, it can be used as a switchable solar absorber in solar collectors. Such
absorbers can be temperature controlled, which make them also ideally suited
for integration with photovoltaic cells.

The authors are grateful to A. Borgschulte, S. Enache, I.A.M.E. Giebels and
A.C. Lokhorst for very useful discussions. Support from the Dutch Stichting
voor Fundamenteel Onderzoek der Materie (FOM) and the Stichting Technische
Wetenschappen (STW) is also acknowledged.


\begin{references}
\bibitem{note1}  Measured values include the Pd cap layer. Therefore,  measured $\rho $ and T are lower than those of the Mg$_{2}$NiH$_{4}$ film itself.
\bibitem{Crystal} Only films with Mg/Ni$>$2 have shown x-ray diffraction peaks.
\end{references}
\end{document}